\documentclass[aps,prd,showpacs,superscriptaddress,preprintnumbers,notitlepage]{revtex4-2}
\usepackage{graphicx,float,wrapfig,subfigure}
\usepackage{amsfonts,amsmath,amssymb,amstext}
\usepackage{latexsym}
\usepackage{bm}
\usepackage{color}
\usepackage[normalem]{ulem}
\usepackage{MnSymbol}
\usepackage[colorlinks=true,linktocpage=true,linkcolor=blue,citecolor=blue]{hyperref}
\usepackage{orcidlink}

\definecolor{red}{rgb}{0.7,0,0}
\definecolor{green}{rgb}{0,0.5,0}

\begin{document} 

\title{Magnetized bottom-up thermalization in heavy-ion collisions}  


\author{Ritesh Ghosh\orcidlink{0000-0002-6740-7038}} 
\email{riteshghosh283@gmail.com} 
\affiliation{Institute of Physics, Academia Sinica, Taipei 11529, Taiwan}
 \affiliation{Institut f\"{u}r Kernphysik, Technische Universit\"{a}t Darmstadt, Schlossgartenstra\ss{}e 2, D-64289 Darmstadt, Germany}

\author{Igor A. Shovkovy\orcidlink{0000-0002-5230-6891}} 
\email{igor.shovkovy@asu.edu} 
\affiliation{College of Integrative Sciences and Arts, Arizona State University, Mesa, Arizona 85212, USA} 
\affiliation{Department of Physics, Arizona State University, Tempe, Arizona 85287, USA}  

\date{\today}  

\begin{abstract}
We investigate how a strong magnetic field generated in noncentral heavy-ion collisions may modify the bottom-up equilibration scenario. In the conventional weak-coupling picture, the earliest stages of the evolution are dominated by overoccupied gluons, while quark production is parametrically delayed. In a background magnetic field, however, additional inelastic channels become kinematically allowed or enhanced, most notably gluon decay into quark-antiquark pairs, $g\to q+\bar q$. Using parametric estimates, we show that for sufficiently strong fields, with $|eB|$ approaching the saturation scale squared, $Q_s^2$, magnetic-field-induced quark production can become important during the earliest stages of bottom-up evolution. This mechanism can populate the hard quark sector, modify the chemical composition of the preequilibrium matter, and provide an additional pathway toward chemical equilibration. We also discuss possible backreaction effects, including quark-antiquark annihilation, depletion of the hard-gluon sector, and the potential feedback of early quark production on the electromagnetic conductivity of the medium. This exploratory study of a magnetically assisted bottom-up scenario provides a natural extension of the standard framework, with qualitative predictions that depend sensitively on the lifetime and spacetime profile of the magnetic field.
\end{abstract}

\maketitle  

\section{Introduction}
\label{Introduction}

The bottom-up equilibration scenario provides a compelling weak-coupling framework to describe how initially overoccupied gluonic matter evolves toward a thermalized quark-gluon plasma (QGP) during the early stages of high-energy heavy-ion collisions \cite{Baier:2000sb}. In this picture, hard gluon modes are diluted by longitudinal expansion while radiating softer gluons, which eventually form a thermal bath. Subsequent energy transfer from the hard sector to this bath drives the system toward equilibration. This multistage mechanism captures essential features of non-Abelian plasma kinetics and has been further refined by kinetic-theory simulations that include anisotropies, plasma instabilities, and collective effects \cite{Shuryak:1992wc,Mrowczynski:1993qm,Arnold:2002zm,Romatschke:2003ms,Lin:2004en,Rebhan:2004ur,Mueller:2005un,Chesler:2008hg,Heller:2011ju,Caceres:2012px,Keegan:2015avk,Kurkela:2018oqw,Ghosh:2020sng}. For reviews, see Refs.~\cite{Arnold:2003rq,Gelis:2015gza,Mrowczynski:2016etf,Busza:2018rrf,Schlichting:2019abc}.

An important ingredient that is often absent from bottom-up equilibration models is the ultrastrong electromagnetic field generated in noncentral heavy-ion collisions \cite{Skokov:2009qp,Voronyuk:2011jd,Deng:2012pc,Bloczynski:2012en,Tuchin:2013ie,Huang:2015oca}. In particular, magnetic fields can reach values of order $|eB|\simeq 10m_\pi^2$ at Relativistic Heavy Ion Collider (RHIC) energies and $|eB|\simeq 70m_\pi^2$ at Large Hadron Collider (LHC) energies during the earliest moments after the collision. Although such fields are short lived, their strength may be sufficient to modify microscopic scattering and splitting processes. As a result, they can introduce qualitatively new inelastic channels and potentially alter the standard picture of bottom-up thermalization.

A background magnetic field enables processes that are either forbidden or strongly suppressed in the conventional bottom-up framework. For example, gluon decay into a quark-antiquark pair, $g\rightarrow q+\bar{q}$, and quark-gluon splitting, $q\rightarrow q+g$, are kinematically forbidden in vacuum by energy-momentum conservation for on-shell particles. In a magnetized medium, however, the transverse motion of charged quarks is quantized, and the external field can absorb transverse momentum. Consequently, such one-to-two and two-to-one processes become kinematically allowed~\cite{Erber:1966vv,Wang:2020dsr,Wang:2021eud,Chen:2024xjj,Ghosh:2024hbf}. These additional inelastic channels provide a mechanism for producing fermionic degrees of freedom at early times and may accelerate chemical equilibration, which is otherwise delayed in the conventional bottom-up scenario. More generally, inelastic processes are also expected to enhance longitudinal momentum diffusion and thereby facilitate kinetic equilibration \cite{Xu:2004mz}.

Incorporating magnetic-field-induced inelastic processes into the bottom-up equilibration scenario is therefore relevant both for understanding QCD matter in extreme background fields and for interpreting phenomenological signatures of the early-time dynamics in heavy-ion collisions. 
To our knowledge, the impact of magnetic fields on the thermalization of strongly coupled plasmas far from equilibrium has been studied only within holographic approaches in Refs.~\cite{Mamo:2015aia,Cartwright:2019opv,Fuini:2015hba}. The conclusions of these studies, however, are not in full agreement: some indicate that magnetic fields can accelerate thermalization~\cite{Mamo:2015aia,Cartwright:2019opv}, while others find only a weak dependence of equilibration times on the external field~\cite{Fuini:2015hba}.

The present study provides an exploratory analysis of magnetically enhanced quark production and its possible role in bottom-up equilibration. Our goal is to identify the parametric regimes in which magnetic effects can compete with, or modify, the standard weak-coupling equilibration mechanisms, and to clarify their potential implications for thermalization and chemical equilibration.

The paper is organized as follows. In Sec.~\ref{bottom-up-m}, we review the basic elements of the bottom-up thermalization scenario and outline how they are modified in the presence of a strong background magnetic field. In Sec.~\ref{bottom-up-q}, we present a qualitative analysis of the magnetized bottom-up framework and derive several predictions for the corresponding equilibration dynamics. In Sec.~\ref{sec:limitations}, we discuss the limitations of the proposed scenario and identify the main issues that should be addressed in future refinements. Finally, Sec.~\ref{summary} summarizes our main findings and offers concluding remarks on magnetically assisted thermalization.

\section{Bottom-up thermalization}
\label{bottom-up-m}

In the bottom-up thermalization scenario, originally developed in Ref.~\cite{Baier:2000sb}, the early-time dynamics of the QGP produced in high-energy heavy-ion collisions are described within weakly coupled quantum chromodynamics. Although this framework is an idealization of the more strongly coupled system likely formed in realistic collisions, it remains conceptually useful and may capture important qualitative features of QGP formation from an initially far-from-equilibrium state.

The bottom-up evolution begins with an overoccupied gluonic system characterized by typical transverse momenta of the order of the saturation scale, $Q_s$. The gluon occupation numbers are parametrically large, $f(p)\sim 1/\alpha_s$, where $\alpha_s\ll 1$ is the strong coupling constant. Thermalization then proceeds through several dynamical stages controlled by the interplay of longitudinal expansion, elastic scatterings, momentum broadening, and inelastic radiation.

In the first stage, extending from $\tau \sim Q_s^{-1}$ to $\tau \sim \alpha_s^{-3/2}Q_s^{-1}$, hard gluons are diluted by longitudinal expansion, while undergoing elastic scatterings and radiating softer gluons. Because the expansion is rapid, the momentum distribution becomes increasingly anisotropic, with longitudinal momenta redshifting relative to transverse momenta. In the second stage, lasting until $\tau \sim \alpha_s^{-5/2}Q_s^{-1}$, the radiated soft gluons accumulate and begin to form a nearly thermal bath with a time-dependent temperature $T_{\text{soft}}(\tau)\sim \alpha_s^2 Q_s^2 \tau$. During this period, inelastic processes, including effective $1\leftrightarrow 2$ splittings and mergings, become increasingly important. In the final stage, the remaining hard gluons lose energy predominantly through medium-induced radiation into the soft bath and experience continued momentum broadening. Parametrically, full thermal equilibrium is reached at 
$\tau_{\text{th}}\sim \alpha_s^{-13/5}Q_s^{-1}$, with the system approaching an isotropic Bose-Einstein distribution characterized by a final temperature $T_{\text{final}}\sim Q_s\alpha_s^{1/4}$.

This scaling picture provides a useful foundation for kinetic-theory descriptions of QGP equilibration in the weak-coupling regime. Its dynamics, however, can be modified by the strong magnetic fields generated in noncentral heavy-ion collisions \cite{Skokov:2009qp,Voronyuk:2011jd,Deng:2012pc,Bloczynski:2012en,Tuchin:2013ie}. During the earliest stages, such strong fields can affect microscopic scattering rates, momentum anisotropies, transport properties, and the relative importance of inelastic channels. As a result, they may alter not only the quantitative timescales but also some qualitative aspects of the equilibration dynamics.

A particularly important effect is the opening of magnetic-field-induced channels for quark production. In the most optimistic estimates, the early-time magnetic field can reach values as large as $|eB|\sim 70m_\pi^2$, which may be comparable to $Q_s^2$. Since the corresponding gluon decay rate into quark-antiquark pairs scales parametrically as $\Gamma_{g\to q\bar q}\sim \alpha_s |eB|/Q_s$ \cite{Ghosh:2024hbf}, this channel can become significant already during the first stage of bottom-up evolution. It leads to the production of hard quarks, modifies the composition of the preequilibrium matter, and provides an additional early-time pathway toward chemical equilibration that is absent in the standard bottom-up framework. While the quantitative importance of this mechanism depends on the lifetime and spacetime profile of the magnetic field, its parametric strength suggests that it should be included in a complete description of weak-coupling equilibration in noncentral collisions. In the next section, we estimate the size of these effects and discuss how they may modify the standard bottom-up scenario.

\section{Magnetized bottom-up scenario}
\label{bottom-up-q}

In the standard scenario, the system created immediately after the heavy-ion collision is dominated by hard gluons with typical transverse momentum of the order of the saturation scale $Q_s$. This scale characterizes the onset of nonlinear gluon interactions in the high-density regime described by the color glass condensate (CGC)~\cite{Gelis:2010nm}. The number of these hard gluons per unit rapidity is approximately conserved during the early stage of the evolution and is given by \cite{Schlichting:2019abc}
\begin{eqnarray}
	\frac{dN}{dy} \sim \frac{\pi}{\alpha_s} Q_s^2 R_A^2,
    \label{dn-dy}
\end{eqnarray}
where $R_A$ is the nuclear radius and $\alpha_s$ is the strong coupling constant. This estimate comes from the CGC framework, where the transverse gluon density scales as $Q_s^2 / \alpha_s$, and the total transverse area of the collision is $\pi R_A^2$.

\subsection{First stage: Early thermalization, $1\lesssim Q_s \tau \lesssim \alpha_s^{-3/2}$}

After the collision, the produced gluonic system undergoes approximately boost-invariant longitudinal expansion along the beam axis. In the Bjorken picture, the longitudinal size of a rapidity slice grows as $\Delta z\sim \tau$, while the transverse area remains approximately fixed, $A_\perp\sim \pi R_A^2$. Thus, the effective volume per unit rapidity scales as
\begin{eqnarray}
    V(\tau) \sim \pi R_A^2 \tau .
\end{eqnarray}
During the early free-streaming stage, the number of hard gluons per unit rapidity is approximately conserved. Therefore, their number density decreases as
\begin{eqnarray}
    n_{g,\text{hard}}(\tau)\sim \frac{1}{V(\tau)}\frac{dN_g}{dy}
    \sim \frac{Q_s^2 R_A^2}{\alpha_s R_A^2 \tau} \sim \frac{Q_s^2}{\alpha_s \tau}.
    \label{n-hard-gluon}
\end{eqnarray}
This $1/\tau$ decrease reflects dilution due to one-dimensional longitudinal expansion. Although the total number of hard gluons per unit rapidity is approximately conserved at this stage, the increasing longitudinal volume reduces their local density.

The early-time system is dominated by hard gluons with typical momentum of order $Q_s$. Its evolution is controlled by the interplay of longitudinal expansion, elastic scatterings, momentum broadening, and inelastic splittings. In the first stage of the bottom-up scenario, with the parametric estimate for the hard-gluon density in Eq.~(\ref{n-hard-gluon}), the corresponding occupation number reads \cite{Baier:2000sb,Cabodevila:2023htm,BarreraCabodevila:2025vir,Schlichting:2019abc,Blaizot:2014jna,BarreraCabodevila:2022jhi,Kurkela:2011ti}
\begin{eqnarray}
    f_{g,\text{hard}} &\sim& \frac{n_{g,\text{hard}}}{p_\perp^2 p_z}\sim \frac{1}{\alpha_s (Q_s \tau)^{2/3}} .
\end{eqnarray}
The typical longitudinal and transverse momenta of hard gluons scale as $ p_z \sim Q_s (Q_s\tau)^{-1/3}$ and $ p_\perp \sim Q_s$, respectively.
Hard gluons also radiate softer gluons through effective $1\to 2$ processes. The density and occupation number of the soft gluon sector are estimated as
\begin{eqnarray}
    n_{g,\text{soft}} &\sim&  \alpha_s \sqrt{\frac{\hat q}{p_s}}\, n_{g,\text{hard}} f_{g,\text{hard}}\,\tau
    \sim \frac{Q_s^3}{\alpha_s (Q_s\tau)^{4/3}}, \\
    f_{g,\text{soft}} &\sim& \frac{n_{g,\text{soft}}}{p_s^3} \sim \frac{1}{\alpha_s (Q_s\tau)^{1/3}} .
\end{eqnarray}
Here $\hat q$ is the transverse-momentum diffusion coefficient, whose parametric estimate in this stage is $\hat q \sim Q_s^3 (Q_s\tau)^{-5/3}$.
The soft gluon distribution can be characterized by an effective temperature $T_{*}$ up to momenta of order $p_s\sim p_z$, with
\begin{eqnarray}
    T_{*} &\sim&  \frac{\hat q}{\alpha_s m_D^2} \sim \frac{Q_s}{\alpha_s (Q_s\tau)^{2/3}},   \\
    m_D^2 &\sim&  \alpha_s \frac{n_{g,\text{hard}}}{Q_s} \sim  \frac{Q_s}{\tau}.
\end{eqnarray}
In the presence of a strong magnetic field, this conventional picture should be amended by including magnetic-field-induced quark production through the process $g\to q+\bar q$. For a hard gluon with energy of order $Q_s$, the corresponding decay rate scales parametrically as
\begin{eqnarray}
    \Gamma_{g\to q\bar q} \sim  \alpha_s \frac{|eB|}{Q_s}.
    \label{Gamma-B-hard}
\end{eqnarray}
This estimate is guided by first-principles studies of photon emission and quark damping in a strong magnetic field \cite{Wang:2020dsr,Wang:2021eud,Chen:2024xjj,Ghosh:2024hbf}. The underlying rates receive contributions from one-to-two and two-to-one processes made kinematically possible by the field. Gluon decay into a quark-antiquark pair is analogous to the inverse of photon production through quark-antiquark annihilation. We therefore expect a similar parametric dependence on the magnetic field and the emitted gauge boson's momentum, with the electromagnetic coupling $\alpha$ replaced by the strong coupling $\alpha_s$. Applying this estimate to an evolving magnetic field assumes that the field varies slowly over the relevant formation time and length. If it does not, the field's variation must be included in the rate calculation. Unfortunately, such a calculation is not currently available and lies beyond the scope of this exploratory study.

If $n_q$ denotes the density of quarks of one charge species, with $n_q\simeq n_{\bar q}$, its evolution in a longitudinally expanding system is more properly written as
\begin{eqnarray}
    \frac{d n_{q,\text{hard}}}{d\tau}+\frac{n_{q,\text{hard}}}{\tau} \simeq \Gamma_{g\to q\bar q}\, n_{g,\text{hard}},
    \label{eq:nq-no-backreaction}
\end{eqnarray}
where the second term on the left-hand side accounts for dilution due to longitudinal expansion. Using $n_{g,\text{hard}}\sim Q_s^2/(\alpha_s\tau)$ and assuming a slowly varying magnetic field, this gives
\begin{eqnarray}
    n_{q,\text{hard}}(\tau)   \sim  |eB|Q_s \left(1-\frac{\tau_0}{\tau}\right),
    \label{eq:nq-first-stage}
\end{eqnarray}
for the initial condition $n_{q,\text{hard}}(\tau_0)=0$. Thus, after a short transient period, the hard-quark density produced by magnetic-field-induced gluon decay is parametrically of order $|eB|Q_s$. The corresponding occupation number is
\begin{eqnarray}
    f_{q,\text{hard}}  \sim \frac{n_q}{Q_s^3} \sim  \frac{|eB|}{Q_s^2}.
\end{eqnarray}
Therefore, for $|eB|\ll Q_s^2$, the produced quark population remains parametrically small and does not strongly affect the gluon sector during the first stage. In contrast, when $|eB|\sim Q_s^2$, the quark occupation number can become of order unity, suggesting that magnetic-field-induced quark production may substantially modify the composition of the preequilibrium matter. In this study, we focus on this stronger-field regime as an optimistic but still plausible early-time scenario. Its relevance to heavy-ion collisions can ultimately be tested by comparing theoretical predictions with experimental observations.

At the end of the first stage, $Q_s\tau\sim \alpha_s^{-3/2}$, the hard-gluon occupation number also becomes of order unity,
\begin{eqnarray}
    f_{g,\text{hard}}   \sim   \frac{1}{\alpha_s (Q_s\tau)^{2/3}} \sim 1 .
\end{eqnarray}
However, the estimate above neglects the inverse process, namely quark-antiquark annihilation into gluons. Without such a back reaction, the quark abundance would continue to grow as long as the magnetic-field-induced source remains active, eventually leading to an unphysical oversaturation of the quark sector. A simple phenomenological way to incorporate microscopic reversibility is to supplement the source term by an annihilation term and write
\begin{eqnarray}
    \frac{d n_{q,\text{hard}}}{d\tau} +\frac{n_{q,\text{hard}}}{\tau}
    \simeq    \Gamma_{g\to q\bar q}  \left( n_{g,\text{hard}} - \frac{n_{q,\text{hard}}^2}{n_{g,\text{hard}}} \right),
    \label{dnq}
\end{eqnarray}
where the second term on the right-hand side describes the annihilation process, $q+\bar q\to g$. In general, the annihilation rate is proportional to $n_q n_{\bar q}$, reflecting the probability for a quark and an antiquark to encounter each other. For a charge-symmetric system, $n_q\simeq n_{\bar q}$, this reduces to a dependence proportional to $n_q^2$ used in Eq.~(\ref{dnq}). The normalization of the inverse term is chosen so that production and annihilation balance parametrically when $n_{q,\text{hard}}\sim n_{g,\text{hard}}$, corresponding to chemical saturation of the hard sector. The model captures both the expected density dependence of the processes and the main qualitative feature of detailed balance: gluon decay dominates when gluons are abundant, while the inverse reaction becomes more important as the quark population grows. It does not, however, enforce detailed balance quantitatively. Doing so would require kinetic collision integrals that account for the momentum distributions, quantum-statistical factors, and matrix elements of both reactions. Thus, Eqs.~(14) and (15) should be understood as phenomenological rate equations.

The corresponding equation for the hard-gluon density can be written as
\begin{eqnarray}
    \frac{d n_{g,\text{hard}}}{d\tau} +\frac{n_{g,\text{hard}}}{\tau}
    \simeq - \Gamma_{g\to q\bar q}  \left( n_{g,\text{hard}} - \frac{n_{q,\text{hard}}^2}{n_{g,\text{hard}}}  \right).
    \label{dng}
\end{eqnarray}
\begin{figure}[b]
	\begin{center}
		\includegraphics[scale=1.]{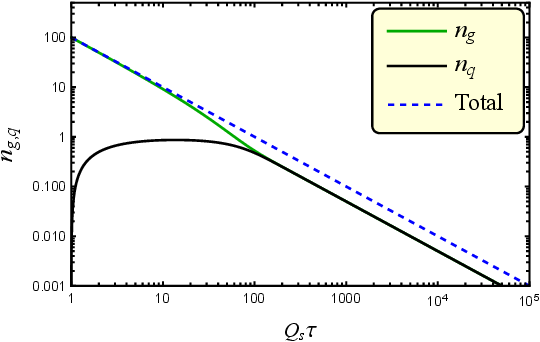}
		\caption{Time evolution of the number densities of hard gluons and hard quarks, together with the total hard-particle density relevant for thermalization.}
		\label{fig:nqg}
	\end{center}
\end{figure}
In the absence of conversion processes, this equation reproduces the Bjorken dilution of the hard-gluon density, $n_{g,\text{hard}}\propto 1/\tau$. Solving Eqs.~(\ref{dnq}) and (\ref{dng}) allows one to track the coupled evolution of hard quark and gluon densities in the presence of both magnetic-field-induced production and inverse annihilation. Our treatment does not describe how energy is redistributed among the particles, but the rate estimates are guided by microscopic calculations in Refs.~\cite{Wang:2020dsr,Wang:2021eud,Chen:2024xjj,Ghosh:2024hbf}, which account for the kinematics, energy thresholds, and energy conservation of the underlying processes. Thus, although these constraints are not explicitly enforced in the present model, they are indirectly incorporated into our order-of-magnitude rates. A fully self-consistent treatment of energy conservation, momentum broadening, and detailed balance would require coupled kinetic equations for the quark and gluon distribution functions.

Our numerical estimates of the hard quark and gluon number densities are shown in Fig.~\ref{fig:nqg} for $\alpha_s = 0.01$. The quark density grows with time and eventually becomes comparable to the gluon density around $\tau \sim Q_s/(\alpha_s |eB|)$. Thus, increasing $\alpha_s$ and/or the magnetic field strength shifts the convergence of the two densities to earlier times, while decreasing either parameter delays it.

The magnetic-field assisted quark production alone does not guarantee much faster kinetic equilibration, isotropization, or hydrodynamization. Nevertheless, by analogy with other inelastic processes known to enhance equilibration \cite{Xu:2004mz}, we suggest that it may accelerate both chemical and kinetic equilibration. Furthermore, the magnetic field can efficiently redistribute particle momenta within the reaction plane (perpendicular to the field), potentially contributing to isotropization. Determining whether these effects truly accelerate kinetic equilibration requires a detailed study of coupled quark--gluon dynamics, which is beyond the scope of this work.

We argue that the production of quarks should dramatically increase the electrical conductivity of the early-time plasma. This could, in turn, affect the lifetime of the in-medium magnetic field and phenomena that depend on it, including the chiral magnetic effect. Quantifying this feedback requires further investigation but our estimate in the Appendix~\ref{app:conductivity-growth} is promising.

\subsection{Second stage: Cooling and overcooling of soft sector, $\alpha_s^{-3/2} \lesssim Q_s \tau \lesssim \alpha_s^{-5/2}$}

At later times, when $Q_s\tau \gtrsim \alpha_s^{-3/2}$, the hard-gluon occupation number drops below unity. The system is still primarily governed by longitudinal expansion, and the hard gluons have not yet lost a substantial fraction of their energy. Their number density therefore continues to scale as
\begin{eqnarray}
    n_{g,\text{hard}} \sim \frac{Q_s^2}{\alpha_s \tau}.
\end{eqnarray}
Once $f_{g,\text{hard}}<1$, Bose enhancement in the hard sector is no longer important, and the jet transport coefficient changes to
\begin{eqnarray}
    \hat{q} \sim  \alpha_s^2 n_{g,\text{hard}} \sim \frac{\alpha_s Q_s^2}{\tau}.
\end{eqnarray}
The typical longitudinal momentum generated by momentum diffusion is then
$p_z \sim \sqrt{\hat{q}\tau} \sim \alpha_s^{1/2} Q_s $. 
Soft gluons with characteristic momentum $p_s\sim p_z$ are produced through effective $1\to 2$ splittings. Their number density and occupation number scale as
\begin{eqnarray}
    n_{g,\text{soft}} &\sim& \alpha_s \sqrt{\frac{\hat{q}}{p_s}}\, n_{g,\text{hard}}\tau
    \sim \alpha_s^{1/4} (Q_s \tau)^{-1/2} Q_s^3, \\
    f_{g,\text{soft}} &\sim& \frac{n_{g,\text{soft}}}{p_s^3} \sim \alpha_s^{-5/4} (Q_s \tau)^{-1/2}.
\end{eqnarray}
Although $n_{g,\text{soft}}$ remains smaller than $n_{g,\text{hard}}$ until $Q_s\tau\sim \alpha_s^{-5/2}$, the soft sector dominates Debye screening because of its larger phase-space density. The corresponding Debye mass is
\begin{eqnarray}
    m_D^2 \sim \alpha_s \frac{n_{g,\text{soft}}}{p_s} \sim \alpha_s^{3/4} (Q_s \tau)^{-1/2} Q_s^2 .
\end{eqnarray}
The effective temperature of the soft sector is estimated as
\begin{eqnarray}
    T_{*} \sim \frac{\hat{q}}{\alpha_s m_D^2} \sim \alpha_s^{-3/4} (Q_s \tau)^{-1/2} Q_s .
\end{eqnarray}
This quantity satisfies $n_{g,\text{soft}}\sim T_{*} p_z^2$, which is consistent with a thermal-like soft distribution up to momenta of order $p_z$.

At this stage, magnetic-field-induced quark production can receive contributions from both hard and soft gluons. For gluons with characteristic momentum $p$, the parametric decay rate in a strong magnetic field scales as $\Gamma_{g\to q\bar q}(p) \sim \alpha_s |eB|/p$. Therefore, while the decay rate for hard gluons with $p\sim Q_s$ is given by Eq.~(\ref{Gamma-B-hard}), the corresponding rate for soft gluons with $p_s\sim \alpha_s^{1/2}Q_s$ scales parametrically as
\begin{eqnarray}
    \Gamma_s \sim \alpha_s \frac{|eB|}{p_s} \sim \alpha_s^{1/2}\frac{|eB|}{Q_s}.
\end{eqnarray}
The evolution of the soft-quark density can then be described by a rate equation of the form
\begin{eqnarray}
    \frac{d n_{q,\text{soft}}}{d\tau} + \frac{n_{q,\text{soft}}}{\tau}
    \simeq \Gamma_s n_{g,\text{soft}} - \gamma_s n_{q,\text{soft}}^2,
    \label{eq:nqs-rate}
\end{eqnarray}
where the term $n_{q,\text{soft}}/\tau$ accounts for Bjorken dilution. The first term on the right-hand side describes quark production from the soft-gluon sector, while the second term parametrizes the inverse annihilation process. Assuming a charge-symmetric system with $n_{q,\mathrm{soft}} \simeq n_{\bar q,\mathrm{soft}}$, we take the annihilation term to be proportional to $n_{q,\mathrm{soft}}^2$.
The coefficient $\gamma_s$ can be estimated from detailed-balance considerations. Requiring the net source term in Eq.~(\ref{eq:nqs-rate}) to vanish once the soft sector reaches chemical saturation, we obtain
\begin{eqnarray}
    \gamma_s \sim \Gamma_s \frac{n_{g,\text{soft}}^{\rm eq}}{\left(n_{q,\text{soft}}^{\rm eq}\right)^2} .
\end{eqnarray}
This form should be understood as a parametric implementation of chemical back reaction rather than as a substitute for a full kinetic treatment.

Neglecting the annihilation term for the purpose of a simple estimate, the soft-sector contribution to the quark density satisfies
\begin{eqnarray}
    \frac{d n_{q,\text{soft}}}{d\tau} + \frac{n_{q,\text{soft}}}{\tau} \sim \Gamma_s n_{g,\text{soft}}.
\end{eqnarray}
Using $\Gamma_s\sim \alpha_s^{1/2}|eB|/Q_s$ and 
$n_{g,\text{soft}}\sim \alpha_s^{1/4}(Q_s\tau)^{-1/2}Q_s^3$, one obtains
\begin{eqnarray}
    n_{q,\text{soft}} \sim \frac{1}{\tau} \int^{\tau} d\tau'\, \tau'\Gamma_s n_{g,\text{soft}}(\tau')
    \sim \alpha_s^{3/4}|eB|Q_s (Q_s\tau)^{1/2},
    \label{eq:nqs-estimate}
\end{eqnarray}
up to numerical factors of order unity and weak dependence on the lower limit of integration. This estimate shows that quark production from the soft sector grows with time during the second stage and can become parametrically important if the magnetic field remains sufficiently strong and long lived.
This may appear to be a strong assumption. However, it is partially supported by the qualitative estimate in Appendix~\ref{app:conductivity-growth}, which shows that early quark production can lead to a buildup of electrical conductivity and, in turn, slow the decay of the magnetic field.

\section{Possible limitations of the scenario}
\label{sec:limitations}

The magnetically modified bottom-up scenario discussed here should be viewed as an exploratory study. While the parametric estimates suggest that magnetic-field-induced quark production may affect the early stages of equilibration, several limitations and uncertainties should be kept in mind.

First, the lifetime of the magnetic field is a crucial issue. The estimates based on the instantaneous rate $\Gamma_g \sim \alpha_s |eB|/Q_s$ are meaningful only if the magnetic field persists long enough for the corresponding conversion probability to become sizable. More precisely, allowing for a time-dependent magnetic field, the survival probability of a gluon against magnetic-field-induced conversion satisfies
\begin{equation}
\frac{dP_g(\tau)}{d\tau}=-\Gamma_g(\tau)P_g(\tau),
\qquad
P_g(\tau_0)=1,
\end{equation}
where $\Gamma_g(\tau) \sim \alpha_s\frac{|eB(\tau)|}{Q_s}$. The corresponding cumulative conversion probability is therefore
\begin{equation}
P_{g\to q\bar q}(\tau)=1-\exp\left[-\int_{\tau_0}^{\tau}d\tau'\,\Gamma_g(\tau') \right].
\label{Pgqq-integrated}
\end{equation}
For a small conversion probability, this reduces to the parametric estimate
\begin{equation}
P_{g\to q\bar q} \sim \int_{\tau_0}^{\tau_1}d\tau\, \frac{\alpha_s |eB(\tau)|}{Q_s},
\end{equation}
used in our previous discussion. Thus, the relevant quantity is the time-integrated magnetic-field strength rather than the initial value of the rate alone. If the field remains of order $|eB|\sim Q_s^2$ throughout a substantial fraction of the first bottom-up stage, $\tau\lesssim \alpha_s^{-3/2}Q_s^{-1}$, then the integrated probability can be parametrically large. In contrast, if the field decays on a time scale of order $Q_s^{-1}$, the probability remains only of order $\alpha_s$ and the effect is parametrically small. Thus, the quantitative importance of this mechanism depends sensitively on the lifetime and spacetime profile of the magnetic field. A systematic sensitivity analysis using realistic time-dependent magnetic-field profiles, including the response of the medium, is beyond the scope of the present exploratory study and is left for future work.

Second, the hierarchy of scales is not universal. The estimate $|eB|\sim Q_s^2$ should be viewed as an optimistic early-time possibility rather than a generic condition. The importance of the magnetic-field-induced channel is controlled by the ratio $|eB|/Q_s^2$. For $|eB|\sim Q_s^2$, the gluon decay rate is parametrically $\Gamma_g\sim \alpha_s Q_s$ and may compete with standard bottom-up dynamics, provided the field persists long enough. For weaker fields, $|eB|\ll Q_s^2$, the effect is correspondingly suppressed.

The use of a local rate derived for a static and homogeneous magnetic field~\cite{Erber:1966vv,Wang:2020dsr,Wang:2021eud,Chen:2024xjj,Ghosh:2024hbf} should likewise be treated with caution. In realistic heavy-ion collisions, the magnetic field is both time dependent and spatially inhomogeneous. The local-rate approximation is justified only when the field varies slowly over the formation time and formation length of the relevant splitting process. If the field changes appreciably during formation, the simple rate estimate must be replaced by a nonlocal treatment of the transition probability.

If efficient, the decay channel $g\rightarrow q+\bar{q}$ would not only populate the quark sector but also deplete the hard gluon sector. The resulting backreaction can modify the screening scale, the momentum-broadening rate, soft-gluon production, and possibly the duration of the individual bottom-up stages. In the present exploratory treatment, such feedback effects are included only semiquantitatively. A more rigorous analysis would require a coupled kinetic description of gluons, quarks, and the evolving magnetic field.

The present scenario also neglects other early-time background fields. In particular, strong chromoelectric and chromomagnetic glasma fields can affect particle production, momentum broadening, and isotropization. The electromagnetic mechanism discussed here should therefore eventually be embedded in a broader framework that includes both electromagnetic and color-field backgrounds.

Finally, treating the magnetic field as an external background is appropriate for a first estimate, but it neglects the backreaction of the produced charged particles on the field itself. In a conducting medium, induced currents can modify both the lifetime and the spatial profile of the magnetic field. In Appendix~\ref{app:conductivity-growth}, we estimate how the conductivity builds up during the early stages of the bottom-up scenario. The results suggest that early quark production can slow the decay of the magnetic field by increasing the conductivity. A more systematic and fully self-consistent treatment, however, would require coupling the kinetic evolution of the preequilibrium plasma to Maxwell's equations or, at later times, to an appropriate magnetohydrodynamic description.

Despite these limitations, the parametric strength of the magnetic-field-induced gluon decay channel suggests that it may provide a non-negligible correction to the standard bottom-up picture when the magnetic field is sufficiently strong and long lived. The scenario should therefore be viewed as a controlled extension of the conventional weak-coupling framework, whose quantitative relevance depends primarily on the spacetime evolution of the magnetic field and on its competition with standard QCD equilibration mechanisms.
 
\section{Conclusions}
\label{summary}

In this work, we proposed an extension of the conventional bottom-up thermalization scenario by incorporating the effects of a strong background magnetic field generated in noncentral heavy-ion collisions. In the standard weak-coupling picture, the early preequilibrium system is dominated by overoccupied gluons, while quark production remains delayed and parametrically suppressed. We argued that this picture can be modified in the presence of a sufficiently strong magnetic field, which opens additional inelastic channels involving quarks and gluons.

The main mechanism considered in this study is magnetic-field-induced gluon decay into quark-antiquark pairs. This process is absent in the standard bottom-up scenario but becomes possible in a magnetized medium because the magnetic field modifies the transverse motion of charged quarks. As a result, hard gluons can convert into quark-antiquark pairs already during the earliest stage of the evolution. When the magnetic field is strong enough and persists for a sufficiently long time, this channel can populate the hard quark sector and modify the chemical composition of the preequilibrium matter.

We also considered the role of the soft sector during the later stages of bottom-up evolution. As soft gluons accumulate and begin to dominate screening, they can provide an additional source of quark production in the presence of the magnetic field. Since softer gluons are more sensitive to magnetic-field-induced splitting processes, the soft sector may further enhance the quark abundance during the intermediate stages of equilibration. This suggests that magnetic effects can influence both the early hard sector and the subsequent evolution of the soft bath.

To make the description more physically consistent, we included phenomenological backreaction terms that account for the inverse process of quark-antiquark annihilation into gluons. These terms prevent unphysical oversaturation of the quark sector and allow the system to approach chemical saturation at the level of a parametric rate-equation treatment. We emphasized, however, that this simplified description does not replace a full kinetic analysis. A complete treatment would require momentum-dependent distribution functions, energy conservation, detailed balance, screening effects, and the time-dependent evolution of the magnetic field.

The central conclusion of this work is that magnetic-field-induced quark production may provide an additional early-time pathway toward chemical equilibration in noncentral heavy-ion collisions. The quantitative importance of this mechanism depends not only on the initial strength of the magnetic field but also on its lifetime and spacetime profile. If the magnetic field decays very rapidly, the effect is likely to remain small. If, however, the field is sustained by medium response or other mechanisms over a substantial fraction of the preequilibrium evolution, magnetically induced quark production may noticeably alter the composition and transport properties of the plasma.

Several directions for future work follow naturally from this exploratory analysis. First, the spacetime dependence of the electromagnetic field should be incorporated explicitly, rather than treating the field as static and homogeneous. Second, the feedback of quark production on the electromagnetic conductivity of the medium should be studied self-consistently. This feedback between quark production, conductivity, and magnetic-field evolution could be important for the persistence of magnetic effects.

Finally, early-produced quarks may also affect momentum transport and isotropization. Their interaction with the magnetic field can redistribute momentum and potentially modify pressure anisotropies during the preequilibrium stage. Quantifying this effect requires going beyond number-density estimates and solving kinetic equations with realistic momentum anisotropies, magnetic-field-dependent collision kernels, and spacetime evolution. Such developments would help determine whether magnetically assisted equilibration can have observable consequences, for example, in electromagnetic probes, charge-dependent correlations, jet quenching, or the centrality dependence of hydrodynamization in heavy-ion collisions.
 
\acknowledgments{R.~G. acknowledges support from Academia Sinica through Project No.~AS-CDA-114-M01, an Academia Sinica Postdoctoral Fellowship during the early stages of this work, and from a Postdoctoral Research Fellowship of the Alexander von Humboldt Foundation during its later stages. I.A.S. was supported in part by the U.S. National Science Foundation under Grant Nos.~PHY-2209470 and PHY-2514933.}

\appendix

\section{Conductivity build-up time}
\label{app:conductivity-growth}
In this appendix, we estimate the timescale over which the electrical conductivity builds up due to magnetic-field-induced quark production. The purpose is to clarify the parametric competition between quark production, conductivity growth, and magnetic-field decay.

A useful local measure of the growth time of the conductivity is
\begin{equation}
    \tau_\sigma \equiv \frac{\sigma}{d\sigma/d\tau}.
    \label{eq:tausigma-def}
\end{equation}
This definition is analogous to a local relaxation or growth time. It should not be interpreted as implying exponential growth; rather, it characterizes the instantaneous rate at which the conductivity changes.

In kinetic theory, electrical conductivity can be estimated parametrically as
\begin{equation}
    \sigma \sim \alpha \frac{n_q}{\langle p\rangle} \tau_{\text{rel}},
    \label{eq:sigma-kinetic-app}
\end{equation}
where $\alpha$ is the fine structure constant and $n_q$ is the density of electrically charged quasiparticles, $\langle p\rangle$ is their typical momentum, and $\tau_{\text{rel}}$ is a microscopic relaxation time. In the early hard sector, $\langle p\rangle\sim Q_s$. If the dominant time dependence of $\sigma$ comes from the quark density, while $\langle p\rangle$ and $\tau_{\text{rel}}$ vary more slowly, then
$\sigma(\tau)\propto n_q(\tau)$, and the conductivity build-up time is estimated as
\begin{equation}
    \tau_\sigma \sim \frac{n_q}{dn_q/d\tau}.
    \label{eq:tausigma-nq}
\end{equation}
The quark density is generated by the magnetic-field-induced inelastic pair-creation process $g\rightarrow q+\bar{q}$. For hard gluons with typical momentum of order $Q_s$ and density given in Eq.~(\ref{n-hard-gluon}), the corresponding rate is given by Eq.~(\ref{Gamma-B-hard}). In an expanding system, the local quark density satisfies
\begin{equation}
    \frac{dn_q}{d\tau} + \frac{n_q}{\tau} \sim \Gamma_{g\to q\bar q} \, n_{g,\text{hard}}.
    \label{eq:nq-expanding-app}
\end{equation}
which is the early-time version of Eq.~(\ref{dnq}), valid when the backreaction remains negligible. Assuming that the magnetic field varies slowly with time and using the parametric estimates given above, Eq.~(\ref{eq:nq-expanding-app}) implies
\begin{equation}
    n_q(\tau) \sim |eB|Q_s \left(1-\frac{\tau_0}{\tau} \right),
    \label{eq:nq-app}
\end{equation}
for the initial condition $n_q(\tau_0)=0$, up to factors of order unity. Thus the local hard-quark density approaches a value of order $|eB|Q_s$ after a time of order $\tau_0$.

Using Eq.~(\ref{eq:nq-app}), the local conductivity build-up time is parametrically
\begin{equation}
    \tau_\sigma \sim \frac{n_q}{dn_q/d\tau} \sim \tau\left(\frac{\tau}{\tau_0}-1\right),
    \label{eq:tausigma-app}
\end{equation}
when only the explicit growth of $n_q$ is used in the denominator. At early times, $\tau-\tau_0\ll \tau_0$, this timescale is of order $\tau-\tau_0$, reflecting rapid initial growth. At later times, the local density saturates in this simplified estimate and $dn_q/d\tau$ becomes small. In that regime, further growth of the conductivity depends on the time dependence of the magnetic field, the relaxation time, the typical particle momentum, and possible contributions from the soft sector.

The magnetic-field decay time in a conducting medium is governed by magnetic diffusion, 
\begin{equation}
    \tau_B\sim \sigma R_{A}^2 \sim \alpha |eB| \left(1-\frac{\tau_0}{\tau} \right)\tau_{\text{rel}} R_{A}^2 ,
    \label{tau-B}
\end{equation}
where we used Eqs.~(\ref{eq:sigma-kinetic-app}) and (\ref{eq:nq-app}). 

The relevant question is therefore whether the conductivity generated by early quark production becomes large before the magnetic field decays. Parametrically, this requires the conductivity build-up time to be shorter than, or at least comparable to, the magnetic diffusion time, i.e., $\tau_\sigma \lesssim \tau_B$. Using the estimates in Eqs.~(\ref{eq:tausigma-app}) and (\ref{tau-B}), and assuming $\tau_{\text{rel}}\sim Q_s/(\alpha_s |eB|)$, this condition translates into 
\begin{equation}
   1 \lesssim \frac{\alpha}{\alpha_s} Q_s^2 R_A^2 \sim \alpha \frac{dN}{dy},
\end{equation}
where we took into account Eq.~(\ref{dn-dy}).
If this condition is satisfied, as may indeed be the case, early quark production can feed back on the electromagnetic evolution by increasing the conductivity and thereby slowing the decay of the magnetic field. Otherwise, the magnetic field decays before the induced conductivity becomes dynamically important.

This qualitative estimate is encouraging. A more quantitative analysis, which lies beyond the scope of the present study, would require a microscopic calculation of the conductivity in a nonequilibrium plasma using kinetic equations for quarks and gluons self-consistently coupled to electromagnetic fields.


%

\end{document}